\documentclass[sigconf]{acmart}
\usepackage{graphicx}
\usepackage{textcomp}

\DeclareMathOperator*{\argmin}{argmin}
\usepackage{multirow}
\usepackage{threeparttable}
\usepackage{algorithm}
\usepackage{algorithmic}
\usepackage{array}
\usepackage{caption} 

\AtBeginDocument{%
  \providecommand\BibTeX{{%
    \normalfont B\kern-0.5em{\scshape i\kern-0.25em b}\kern-0.8em\TeX}}}

\copyrightyear{2023}
\acmYear{2023}
\setcopyright{acmlicensed}\acmConference[SIGSPATIAL '23]{The 31st ACM International Conference on Advances in Geographic Information Systems}{November 13--16, 2023}{Hamburg, Germany}
\acmBooktitle{The 31st ACM International Conference on Advances in Geographic Information Systems (SIGSPATIAL '23), November 13--16, 2023, Hamburg, Germany}
\acmPrice{15.00}
\acmDOI{10.1145/3589132.3625582}
\acmISBN{979-8-4007-0168-9/23/11}

\begin{document}

\title{FedGeo: Privacy-Preserving User Next Location Prediction \\ with Federated Learning}



\author{Chung Park}

\email{cpark88kr@gmail.com}
\affiliation{%
  \institution{SK Telecom, KAIST}
  \country{Republic of Korea}
}

\author{Taekyoon Choi}
\email{tgchoi03@gmail.com}
\affiliation{%
  \institution{NAVER}
  \country{Republic of Korea}}

\author{Taesan Kim}
\email{ktmountain@sk.com}
\affiliation{%
  \institution{SK Telecom}
  \country{Republic of Korea}
}

\author{Mincheol Cho}
 \email{skt.mccho@sk.com}
\affiliation{%
 \institution{SK Telecom}
 \country{Republic of Korea}}

\author{Junui Hong}
\email{skt.juhong@sk.com}
\affiliation{%
  \institution{SK Telecom, KAIST}
  \country{Republic of Korea}}

\author{Minsung Choi}
\email{ms.choi@sk.com}
\affiliation{%
  \institution{SK Telecom}
  \country{Republic of Korea}
  }

\author{Jaegul Choo}
\authornote{Corresponding Author (Korea Advanced Institute of Science and Technology)}
\email{jchoo@kaist.ac.kr}
\affiliation{%
  \institution{KAIST}
  \country{Republic of Korea}}

\renewcommand{\shortauthors}{Chung Park et al.}

\begin{abstract}
A User Next Location Prediction (UNLP) task, which predicts the next location that a user will move to given his/her trajectory, is an indispensable task for a wide range of applications.
Previous studies using large-scale trajectory datasets in a single server have achieved remarkable performance in UNLP task. 
However, in real-world applications, legal and ethical issues have been raised regarding privacy concerns leading to restrictions against sharing human trajectory datasets to any other server. 
In response, Federated Learning (FL) has emerged to address the personal privacy issue by collaboratively training multiple clients (i.e., users) and then aggregating them. 
While previous studies employed FL for UNLP, they are still unable to achieve reliable performance because of the heterogeneity of clients' mobility. 
To tackle this problem, we propose the \underline{Fed}erated Learning for \underline{Geo}graphic Information (FedGeo), a FL framework specialized for UNLP, which alleviates the heterogeneity of clients' mobility and guarantees personal privacy protection. 
Firstly, we incorporate prior global geographic adjacency information to the local client model, since the spatial correlation between locations is trained partially in each client who has only a heterogeneous subset of the overall trajectories in FL.
We also introduce a novel aggregation method that minimizes the gap between client models to solve the problem of client drift caused by differences between client models when learning with their heterogeneous data.
Lastly, we probabilistically exclude clients with extremely heterogeneous data from the FL process by focusing on clients who visit relatively diverse locations.
We show that FedGeo is superior to other FL methods for model performance in UNLP task.
We also validated our model in a real-world application using our own customers' mobile phones and the FL agent system.
\end{abstract}





\maketitle

\section{Introduction}
\;\;\;\;\;User Next Location Prediction (UNLP) is one of the most popular uses of mobility mining, and is employed in various applications for points of interest (POI) recommendation, location-based targeting, and mobility pattern modeling \cite{liu2016predicting,lin2020pre,park2023pre}.
The UNLP model infers the next location of the corresponding trajectory, which is a sequence of locations (e.g., GPS coordinates).
Previous studies using large-scale trajectory data in a single server have achieved remarkable performance on the UNLP task.
This learning scheme is centralized learning (CL), which is commonly used in the machine learning field \cite{lin2020pre,zhou2018deepmove}.
However, some legal and ethical issues protecting personal privacy restrict human trajectory data from being transferred to any other server.
Therefore, previous UNLP models are not suitable for real-world applications \cite{hard2018federated,bhowmick2018protection,rodriguez2021enforcing}.

To tackle this issue, we propose a privacy-preserving UNLP model using a federated learning (FL) framework\cite{mcmahan2017communication} in which multiple clients (e.g., user mobile phone) individually train their models under the orchestration of a central server.
In the federated learning, several clients such as mobile phones or IoT devices store training data independently, while the centralized learning framework stores the whole dataset on a server\cite{karimireddy2020scaffold}.
The federated learning process is summarized as follows: 
(1) First, a server samples a fraction of the clients participating in the federated learning.
(2) Then, a global model in the server is sent to the sampled clients.
(3) Each client trains its model initialized by the global model weights with their own dataset.
(4) The trained models in the sampled clients are aggregated using a specific method, such as averaging (i.e., FedAvg \cite{mcmahan2017communication}).
This process updates the global model and is repeated until the global model converges.
Note that there is no transmission of a specific client's dataset to the server and other clients.

\begin{figure}[h]
    \begin{center}
    \includegraphics[width=1\linewidth]{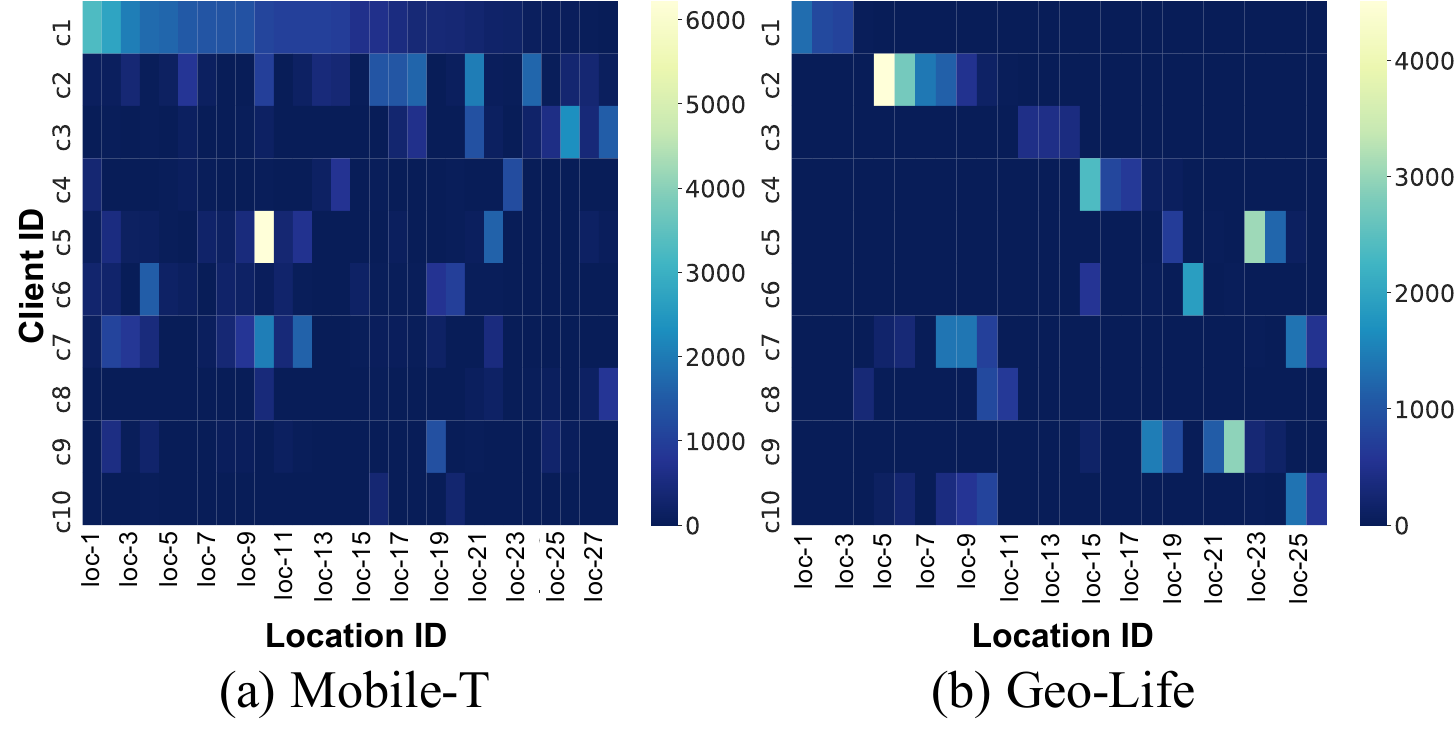}
    \end{center}
    \vspace{-0.6cm}
\caption{Location ID distributions of trajectories in our datasets. 
We illustrate the number of data samples of a specific location ID in each client.
        }
    \label{fig:motivation2}
    \end{figure}

While some previous studies adopted the federated learning for the UNLP task\cite{wang2022location,feng2020pmf,gurukar2021locationtrails}, existing algorithms have yet to achieve reliable performance with deep learning models.
In addition, integrating the UNLP task into existing federated learning methods, such as FedAvg\cite{mcmahan2017communication} or MOON\cite{li2021model}, also proves challenging in achieving stable performance.
This is mainly because the clients' mobility is extremely and statistically heterogeneous\cite{karimireddy2020scaffold}. 
To demonstrate the heterogeneity of user mobility, we illustrate the location distributions of trajectories across users using two datasets used in our experiments, as shown in Figure \ref{fig:motivation2}.
In Figure \ref{fig:motivation2}a, we can see that $\texttt{c5}$ client mostly stays at $\texttt{loc10}$, while $\texttt{c2}$ client in Figure \ref{fig:motivation2}b mostly visits $\texttt{loc5}$ and $\texttt{loc6}$.
The different distributions of frequently visited locations for each user indicate the heterogeneity of each user's mobility dataset.

To circumvent this limitation, we propose the \underline{Fed}erated Learning for \underline{Geo}graphic Information (\textbf{FedGeo}), a simple but novel federated learning framework for the UNLP task, which alleviates the heterogeneity of mobility data and prevents personal privacy leakage.
FedGeo consists of the following three components.

\begin{figure}
    \begin{center}
    \includegraphics[width=0.95\linewidth]{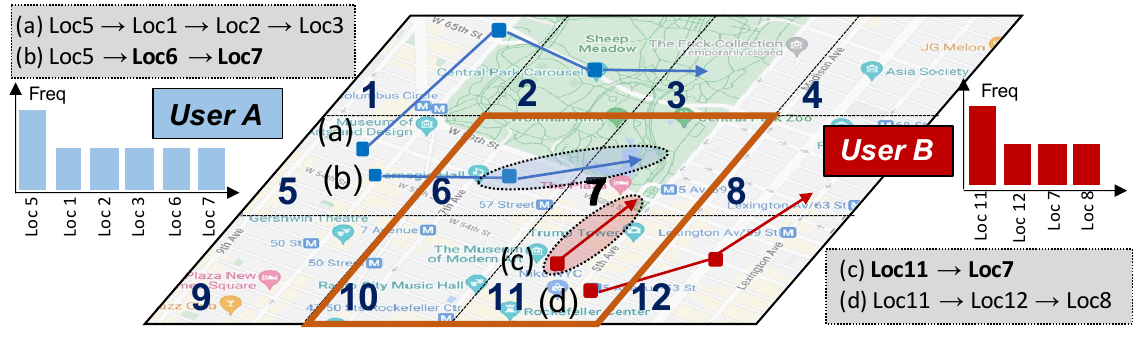} 
    \end{center}
    \vspace{-0.5cm}
\caption{Illustration of trajectories for two users. Numbers on the grid map indicate location IDs. 
In the FL, user \textit{A}'s model can only learn the relationship between \texttt{loc6} and \texttt{loc7}, and user \textit{B}'s model can only learn the relationship between \texttt{loc7} and \texttt{loc11}. GAA provides both models with the information that \texttt{loc7} is geographicly adjacent to \texttt{loc6} and \texttt{loc11}.
}
    \label{fig:motivation}
    \end{figure}

\noindent\textbf{(1) Geographic Adjacency Alignment}: 
The spatial relationship between locations is partially trained in each client who has only a heterogeneous subset of the total trajectories in the federated learning.
As shown in Figure \ref{fig:motivation}, in the federated learning scheme, user \textit{A}'s model can only learn the relationship between \texttt{loc7} and \texttt{loc6}, and user \textit{B}'s model can only learn the relationship between \texttt{loc7} and \texttt{loc11}.
However, \texttt{loc7} has spatial correlation with both \texttt{loc6} and \texttt{loc11}.
Therefore, we propose a Geographic Adjacency Alignment (GAA) that gives \textbf{prior global geographic adjacency information} to the local client models in every round.
This ensures that even if a location is not visited by a particular client, the spatial neighborhood of that location is reflected in the client model.
This module simply multiplies the embedding matrix in the UNLP model by a normalized spatial weight matrix \cite{anselin1995local} representing the geographic adjacency between locations.
Note that the UNLP model is the initialized model just before it is sent to clients in each round.
Using this process, the global geographic adjacency can be proactively reflected in each client model.

\noindent\textbf{(2) Layer-wise Similarity-based Aggregation}:  
The disparity among client models arises from the heterogeneity of user trajectories.
Specifically, the difference between the true global optimum and the average of client updates becomes more prominent as the communication round continues, known as \textit{client drift}\cite{karimireddy2020scaffold}. 
To resolve this problem, previous studies aggregated client models with a euclidean distance between the \textbf{previous} global model in the server and the \textbf{current} client models \cite{ji2019learning,li2018fedprox}; however, restraining the drift with the previous server model induces each client model to gather less novel information\cite{mendieta2022local}.
For this reason, we devised a Layer-wise Similarity-based Aggregation (LWA), which exploits the layer-wise similarity between the \textbf{current} temporary server model aggregated by FedAvg\cite{mcmahan2017communication} and the \textbf{current} client models. 

\noindent\textbf{(3) Entropy-based Sampling}: 
Despite the heterogeneity of user trajectories, clients participating in each round were uniformly sampled in the existing federated learning methods\cite{mcmahan2017communication,karimireddy2020scaffold,ji2019learning,li2018fedprox}.
A client with a balanced distribution of various locations can learn more diverse mobility patterns, thus helping to generalize the server model.
Therefore, we propose an Entropy-based Sampling (EBS) method that selects a set of clients to participate in each round according to the diversity of their mobility patterns.
This method is the sampling mechanism selects clients trained with less heterogeneous dataset.

These three components are incorporated into FedGeo.
The main contributions of this paper are summarized as follows:

\begin{itemize}
  \setlength{\itemsep}{0pt}
  \item We propose FedGeo, a federated learning framework specialized for UNLP, which alleviates the heterogeneity of trajectory data and is robust to the privacy issue.
  \item FedGeo consists of the Geographic Adjacency Alignment to provide a prior geographic adjacency information to local client models, the Layer-wise Similarity-based Aggregation to handle the client drift, and the Entropy-based Sampling to select clients with diverse mobility patterns to frequently participate in the federated learning process.
  \item FedGeo outperformed other federated learning frameworks in terms of model accuracy and performance stability in the UNLP task.
\end{itemize}

\section{Background and Related Work}

\subsection{User Next Location Prediction (UNLP)}

\;\;\;\;\;UNLP is a fundamental task in Location-Based Service (LBS)\cite{liu2016predicting,lin2020pre}.
Previous studies proposed generating location embeddings from sequential trajectories to predict the next location\cite{gao2017identifying, zhou2018deepmove}.
For example, ST-RNN\cite{liu2016predicting} is an RNN-based location embedding model that accounts for both temporal and spatial contexts. It is utilized in predicting the next location.
CTLE\cite{lin2020pre} is a self-attention-based model for constructing a context-aware location embedding model that takes into account the contexts of a target location and has shown the state-of-the-art performance in the UNLP task.

However, previous studies leveraged large-scale trajectory data in a single server for the UNLP task, creating risk and increased responsibility for storing it in a centralized location.
In particular, user trajectory data contains private user information such as GPS coordinates or timestamps; therefore, it is too sensitive and confidential to train a UNLP model using trajectory data stored in the central server.
For this reason, big tech companies such as Apple or Google are constantly working on ways to provide better and more personalized services without violating any privacy restrictions (e.g., Google G-board) \cite{hard2018federated,bhowmick2018protection,rodriguez2021enforcing}.
To address this issue, we propose a privacy-preserving UNLP model applying the federated learning framework\cite{mcmahan2017communication} described in the next section.

\subsection{Federated Learning} \label{ssec:related_work_fl}
\;\;\;\;\;In the federated learning, decentralized clients collaboratively train their models without compromising personal privacy.
The general federated learning process can be summarized as follows: 
(1) First, a server selects a random subset of clients. At this stage, the ratio of client selections in each round is predetermined, called the \textbf{fraction rate}.
(2) Then, the server transmits the global model to the selected clients.
(3) After receiving the global model, each client initializes its local model using the weights of the global model. Subsequently, each client independently trains its local model using its own dataset. At this stage, the number of epochs in each client denotes the \textbf{local epoch}.
(4) The trained models from the sampled clients are aggregated using a particular method, and these (1)-(4) steps are referred to as one \textbf{round}. 
This process updates the global model and is repeated until the global model converges.

There have been several federated learning studies based on aggregation methods or model optimizers.
For example, FedAvg\cite{mcmahan2017communication} calculated the weighted average to parameters of client models with proportional sizes of the client datasets.
FedAdam\cite{reddi2020adaptive} adopted the federated version of adaptive optimizers to tune unfavorable convergence behavior in FedAvg.
While it has been successful in certain applications, there are still some shortcomings with data heterogeneity.
The average of the client models is optimized with different local objectives, causing movement away from the global optimum, known as the \textbf{client drift}.
In recent studies, some research suggested solutions to mitigate the client drift problem.
FedProx\cite{li2018fedprox} added a proximal term to the loss functions in client models by regularizing their gradient updates to remain close to the previous global model.
MOON\cite{li2021model} addressed client drift by using contrastive learning to maximize the agreement of a current client and a previous global model.

Meanwhile, several previous works have studied federated learning applications for UNLP.
For example, LocationTrails\cite{gurukar2021locationtrails} is a skip-gram-based location embedding model under the FedAvg framework that performs a weighted average with client models.
APF \cite{wang2022location} combined a global and a local UNLP models with a certain ratio to derive a personalized UNLP model, using FedAvg for the federated learning framework and the stack of the self-attention layers for the UNLP model.
Similarly, AMF\cite{li2020predicting} adopted the self-attention model for the UNLP model and aggregated the client models with the similarity between the client models and the global model.
PMF\cite{feng2020pmf} proposed the long short-term memroy (LSTM) based UNLP model with the FedAvg scheme. 
They aggregated clients' local models using their training loss in the weighted average instead of their sample size.

However, previous studies remain limited in designing the privacy-preserving UNLP model due to the heterogeneity of user mobility.
For this reason, they showed sub-optimal performance in the UNLP task.
In this paper, we focus on the privacy-preserving UNLP model showing a reliable performance, even with the heterogeneous trajectory dataset in each client.

\section{Preliminaries} \label{sec:Preliminaries}
\textbf{Definition 1. Trajectory}: 
Suppose that $L$ is a set of distinct locations appearing in the training dataset.
The size of $L$ is denoted as $|L|$.
Each location $l$ is represented as a visiting record.
A sequence of locations where a user visited for a specific period is defined as a trajectory\cite{lin2020pre, gao2017identifying, park2023pre}, and can be described as follows,

 \begin{equation}
 \label{equation:trajectory}
  x=\left\{l_{1}, l_{2},\ \ldots, l_{T}\right\}
 \end{equation}
 where $T$ is the length of the trajectory $x$. 
 We also denote $X$ as a set of trajectories of all users.

\noindent\textbf{Definition 2. User Next Location Prediction}: 
Given a trajectory $x=\{l_{1}, l_{2},\ \ldots, l_{T}\}$, a function $f_w$ is to predict the next location $l_{T+1}$, where $w$ are trainable parameters.

Suppose that our UNLP model $f_w$ consists of three parts: (1) an embedding layer, (2) a base encoder, and (3) an output layer\cite{li2021model,park2023pre}.
There are $Z$ layers in the UNLP model $f_w$.
For example, the first layer of $f_w$ is the embedding layer and the last layer is the output layer.
The embedding layer is represented by a matrix $e\in\mathbb{R}^{|L| \times E}$, where $|L|$ is the size of distinct locations and $E$ is the embedding dimension.
The base encoder plays a crucial role in extracting representation vectors to predict the next location from a sequence of location embeddings.
In previous studies, there are various base encoders to predict the next location, such as CTLE\cite{lin2020pre} with the multi-head self-attention layer and ST-RNN\cite{liu2016predicting} with an RNN\cite{hochreiter1997long}.
The output layer is represented by a matrix $h\in\mathbb{R}^{E \times |L|}$.
The output layer produces predicted logits for each class corresponding to the next location.

\noindent\textbf{Problem Setup}: 
We aim to construct a single global UNLP model with the federated learning framework.
Suppose there are \textit{K} clients denoted as $C_1,C_2,...,C_K$ and a central server $S$.
A client $C_k$ represents a specific user and has a local dataset $X_k$.
We set the whole dataset $X$ as the collection of local dataset $X_k$ ($X=\cup_{k\in K}X_{k}$).
We denote $(\textit{x},y)$ as a sample in $X$, where \textit{x} is a trajectory until the $T$ timestamp and $y$ is the corresponding next location $l_{T+1}$.
The UNLP model $f_w$ is shared by all clients with the same model architecture and hyper-parameters setting.
The objective of the federated learning is represented as:

 \begin{equation}
 \begin{split}
 \label{equation:objective}
 w^{*}=\argmin_{w} \mathcal{L}(w)=\argmin_{w} \sum_{k=1}^{K} p_k \mathcal{L}_k(w),\\
 \end{split}
 \end{equation}
where $\mathcal{L}_k(w)=\mathbb{E}_{(\textit{x},y)\sim X_k}[\ell_k(w;(\textit{x},y))]$ is the empirical loss of $C_k$, $\ell_k$ is the cross-entropy loss of the $k$-th client model, and $p_k$ is a parameter to weight client $C_k$.
For FedAvg\cite{mcmahan2017communication}, $p_k$ is described as $\frac{\mid X_k\mid}{\mid X \mid}$.

\section{Model}
\begin{figure*}
    \begin{center}
    \includegraphics[width=0.65\linewidth]{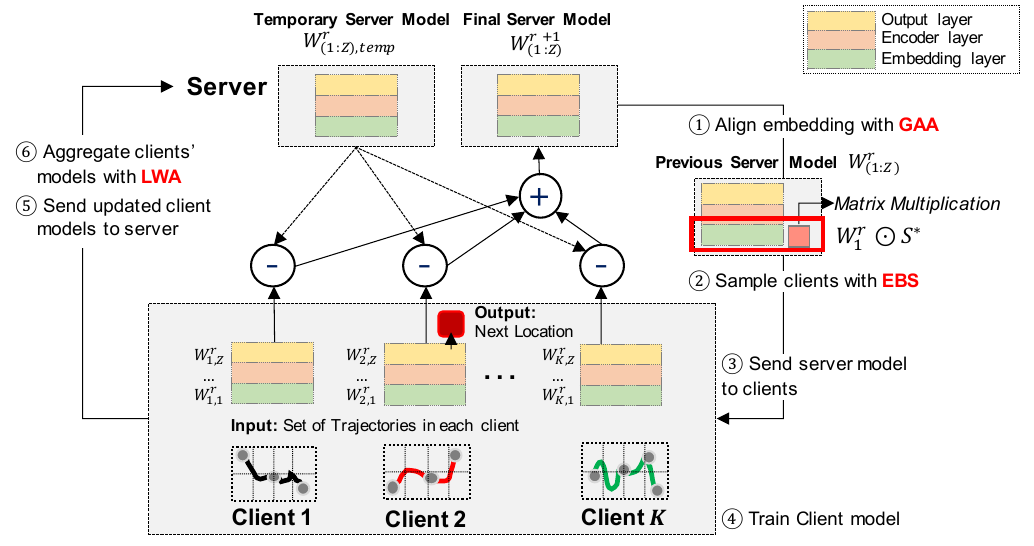}
    \end{center}
    \vspace{-0.3cm}
\caption{Training process of FedGeo. 
Each client has its own set of trajectories and trains the UNLP model independently. 
Note that $\odot$ indicates the matrix multiplication.
We refer to a part of notations in FedAtt\cite{ji2019learning}. 
The notations of $\oplus$ and $\ominus$ represent the layer-wise operation on the weights. 
The notation $\oplus$ indicates the aggregated summation with layer-wise similarities. 
The notation $\ominus$ means a calculation of similarity between the temporary global model $W_{(1:Z),temp}^{r}$ and $K$ client models $W_{(1:Z),k}^{r+1}$.}
    \label{fig:overview}
    \end{figure*}

\;\;\;\;\;Here, we illustrate the process of FedGeo with the UNLP model in Figure \ref{fig:overview}.
In addition, to demonstrate our suggested approach, we summarize the full details in Algorithm \ref{alg:algorithm_fedgeo}. 

    \begin{figure}
    \begin{center}
    \includegraphics[width=1\linewidth]{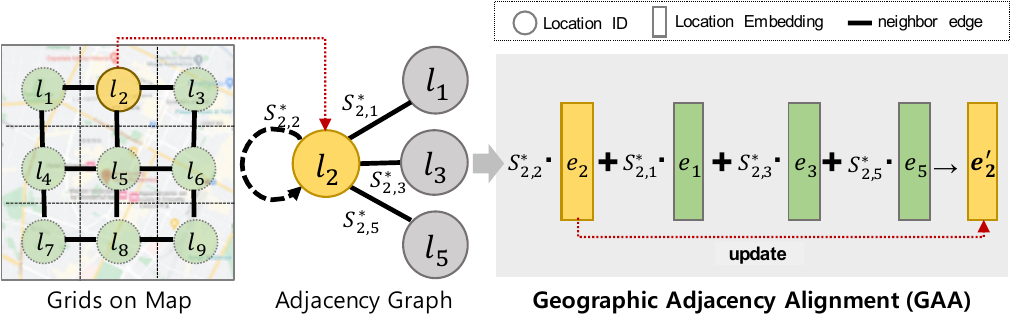}
    \end{center}
    \vspace{-0.3cm}
\caption{We illustrate the operation of geographic Adjacency Alignment. We focus on the location $l_2$, which is adjacent to $l_1$, $l_3$, and $l_5$. 
Embedding vectors for $l_2$, $l_1$, $l_3$, and $l_5$ are denoted as $e_2$, $e_1$, $e_3$, and $e_5$.
The edges between location $i$ and $j$ are described as spatial weight matrix $S_{i,j}^{*}$.
These edges are used as coefficients for the linear combination of embedding vectors $e_2$, $e_1$, $e_3$, and $e_5$.
This operation updates the embedding vector $e_2$ with the neighboring location's embedding vectors.
}
    \label{fig:gaa}
    \end{figure}

\subsection{Geographic Adjacency Alignment}

\;\;\;\;\;In the centralized learning, all user trajectories are used to train the UNLP model; therefore, various mobility patterns are reflected in the model.
In the federated learning, on the other hand, the spatial relationship between locations is partially trained in each client, which has only a heterogeneous subset of all trajectories.
Therefore, we propose the Geographic Adjacency Alignment (GAA), which provides the local client models with prior global geographic adjacency information in each round.

The geographic adjacency is a distinctive characteristic of the trajectory dataset compared to other data, which pre-defines the relationship between locations concerning their geographic distance.
This is significant in the UNLP task since the input trajectory and the corresponding next location cannot be far from each other.
The geographic adjacency between locations can be defined using the location's coordinates in advance, derived from a spatial weight matrix \cite{anselin1995local}.
The spatial weight matrix $W_{i,j}$ represents the degree of the geographic adjacency between location $i$ and $j$.
The element of spatial weight matrix along with location $i$ and $j$ is described as follows:

\begin{equation}
\label{equation:spatial_weight_matrix}
S_{i,j}=\left \{\begin{array}{lll}
      1 \;\;\;\;\;\; if \; i\neq j \;\; and \;\; d_{i,j}<d
     & \\ q \;\;\;\;\;\; if \; i=j
     &  \\ 0 \;\;\;\;\;\; otherwise,
\end{array} \right.
\end{equation}
where $d$ is the threshold determining neighbors and $d_{i,j}$ is the distance between location $i$ and $j$.
The diagonal element is set to $q$, which indicates the self-spatial weight of a particular location $i$.
The spatial weight of neighboring locations is comparatively small when $q$ is large.
As a user can pre-define the value of $q$, therefore we further investigated the effectiveness of $q$ in the experimental section (Section \ref{subsubsection: gaa_q}).
Then, referring the spatial statistics \cite{anselin1995local}, $S_{i,j}$ is standardized with the row axis as follows:

\begin{equation}
\label{equation:standarized_spatial_weight_matrix}
S_{i,j}^{*}=\texttt{Norm}(S_{i,j})=\frac{S_{i,j}}{\sum_{all j} S_{i,j}},
\end{equation}
where \texttt{Norm}$(\cdot)$ denotes the row-normalized function, used for reliable model training.
The standardized spatial weight matrix $S^{*}\in \mathbb{R}^{|L|\times |L|}$ is used to reflect the geographic adjacency to the global model to be sent to each client by multiplying $S^{*}$ by the embedding layer $e\in \mathbb{R}^{|L|\times E}$ of each client model.
Suppose $W_{z}^{r}$ is the $z$-th layer of the global model in the server at the $r$-th round: this operation is described as,

\begin{equation}
\label{equation:multiplication}
W_{1}^{r} \gets S^{*} \cdot W_{1}^{r},
\end{equation}
where $W_{1}^{r}$ is the embedding layer, the first layer of the UNLP model in the $r$-th round.
Note that this operation is applied to the global model on the server just before it is sent to the clients each round.
In this process, a specific location's embedding vector is updated with the averaged vector of the embedding vectors of the neighboring locations.
This allows the embedding of a specific location to proactively reflect the embedding of nearby locations.
The illustration of the geographic adjacency alignment is described in Figure \ref{fig:gaa}.

\begin{algorithm}[tb]
\caption{FedGeo}
\label{alg:algorithm_fedgeo}
\begin{flushleft}
\textbf{Input}: (1) number of clients $K$; (2) number of local epochs $E$;
(3) number of communication rounds $R$; (4) client selection size $G$;
(5) geographic adjacency matrix $S_{i,j}\in\mathbb{R}^{|L|\times |L|}$; (6) set of local dataset $\{X_{1},X_{2},...,X_{K}\}$. \\
\noindent\textbf{Output}: final global model of $Z$-layers $W_{1}^{R}$, $W_{2}^{R}$,...,$W_{Z}^{R}$.
\end{flushleft}

\begin{algorithmic}[1]
\STATE \textbf{SERVER OPERATIONS:}

\STATE Initialize global models of $Z$-layers $W_{1}^{0}$, $W_{2}^{0}$, $...$ ,$W_{Z}^{0}$ ($W_{1}$=embedding layer), and set $\alpha_{z,k}^{0}$ to $0$.
\FOR {each round $r=0,1,...,R-1$}
    \STATE \colorbox{blue!10}{$W_{1}^{r} \gets \texttt{Norm}(S_{i,j})W_{1}^{r}$  \COMMENT {\textcolor{red}{\textbf{\textit{GAA}}}; 
    Eq.\ref{equation:multiplication}}}
    \STATE $K^{r} \gets max(G\cdot K,1)$
    \STATE \colorbox{green!10}{$M^{r} \gets$ set of $K^r$ clients with
$p$ \COMMENT {\textcolor{red}{\textbf{\textit{EBS}}}; Eq.\ref{equation:entropy_sampling}}}
    \STATE Send the global model $W_{1}^{r}$, $W_{2}^{r}$,...,$W_{Z}^{r}$ to set of selected clients $M^r$

        \FOR {each client $k \in M^{r}$ in parallel}
            \STATE $W_{1,k}^{r},...,W_{Z,k}^{r} \gets$ \textbf{Client Update}($W_{1}^{r}$,...,$W_{Z}^{r};X_{k},E$)
        \ENDFOR

\STATE \colorbox{red!10}{$W_{z,temp}^{r} \gets \sum_{k\in M^{r}}\frac{n_k}{n_{M^{r}}}W_{z,k}^{r}$ 
\COMMENT {\textcolor{red}{\textbf{\textit{Temporary Agg}}};  Eq.\ref{equation:layer_wise_agg_1}}}
\STATE \colorbox{red!10}{$\alpha_{z,k}^{r} \gets sim(W_{z,k}^{r}, W_{z,temp}^{r})$ \COMMENT {\textcolor{red}{\textbf{\textit{Similarity}}};  Eq.\ref{equation:dot_product_attention}}} 

\STATE \colorbox{red!10}{$W_{z}^{r+1} \gets \sum_{k\in M^{r}}\alpha_{z,k}^{r}W_{z,k}^{r}$ \COMMENT {\textcolor{red}{\textbf{\textit{Final Agg}}};  Eq.\ref{equation:layer_wise_agg_3}}}

\ENDFOR
\STATE \textbf{Return} $W_{1}^{R}$, $W_{2}^{R}$,...,$W_{Z}^{R}$

\end{algorithmic}
\end{algorithm}

\subsection{Layer-wise Similarity-based Aggregation}
\;\; 
The client drift, which indicates the large gap between client models, varies across layers of the client models \cite{li2018fedprox,ji2019learning}.
For this reason, previous federated learning studies such as FedProx\cite{li2018fedprox}, MOON\cite{li2021model}, and FedAtt\cite{ji2019learning} aggregated client models with the distance between previous global (i.e., round $r$-1) and each current client model (i.e., round $r$) to mitigate the client drift.
However, restricting the drift with the previous global model in the server induces current client models to collect less novel information compared to the global model in the previous round\cite{mendieta2022local}.
For this reason, we propose the Layer-Wise Similarity-based Aggregation (LWA) to aggregate client models with the current layer-wise contribution.
First, the server temporally aggregates client models with the FedAvg scheme at $r$-round as follows:

\begin{equation}
\label{equation:layer_wise_agg_1}
W_{z,temp}^{r} = \sum_{k\in M^{r}}\frac{n_k}{n_{M^{r}}}W_{z,k}^{r},
\end{equation}
where $W_{z,temp}^{r}$ is the weights in the $z$-th layer of the temporary aggregated model at the $r$-th round, $M^{r}$ is a set of clients participating in the $r$ round, and $W_{z,k}^{r}$ is the weights of the $z$-th layer of the $k$-th client at the $r$-th round.
And, $n_{M^{r}}$ is the total number of samples of clients participating in the $r$-round and $n_k$ is the number of data samples of $k$-th clients. 
$W_{z,temp}^{r}$ is the current global model (i.e., round $r$) of the FedAvg method. 
This FedAvg-based global model suffers from the client drift problem.
Therefore, we perform the following additional operations.
We calculate the similarity of the $z$-th layer between the FedAvg-based temporary global model $W_{z,temp}^{r}$ and the $k$-th client model $W_{z,k}^{r}$ as follows:

\begin{equation}
\label{equation:layer_wise_agg_2}
\alpha_{z,k}^{r} = sim(W_{z,k}^{r}, W_{z,temp}^{r}).
\end{equation}
To define the similarity function, we adopt a dot-product method,
\begin{equation}
\label{equation:dot_product_attention}
sim(W_{i},W_{j}) = softmax(\frac{W_{i}W_{j}^{T}}{\sqrt{d_{w}}}).
\end{equation}
Note that we flatten $W_i$ and $W_j$ for the dot-product operation.

Lastly, each client model is aggregated by layer-wise similarity as follows:

\begin{equation}
\label{equation:layer_wise_agg_3}
W_{z}^{r+1} = \sum_{k\in M_{t}}\alpha_{z,k}^{r}W_{z,k}^{r},
\end{equation}
where $W_{z}^{r+1}$ is the $z$-th layer of the final aggregated model for the $(r$+$1)$-round.
The layer-wise similarity $\alpha_{z,k}^{r}$ is the attentive weight to minimize the distance between the global model and the client models.
In short, this operation reduces the degree of the client drift by aggregating client models that are relatively similar to the FedAvg-based global average model with the greater attentive weight.

However, the application of LWA incurs significant computational costs (Eq. \ref{equation:layer_wise_agg_2} and Eq. \ref{equation:layer_wise_agg_3}). 
To mitigate this issue, we examined the selective LWA application to specific layers.
We calculated the similarity values ($\alpha_{z,k}^{r}$) of each layer across all clients based on Eq \ref{equation:layer_wise_agg_2}. 
Our results indicated that the similarity values of the embedding layer and the base encoder were nearly identical among client models, resulting in LWA having a negligible impact. 
Conversely, the output layer exhibited significant differences in similarity across clients. 
Thus, the computational cost can be reduced by selectively applying LWA to specific layers such as the output layer within the model.

\vspace{-0.3cm}
\subsection{Entropy-based Sampling}

\;\;\;\;\;The heterogeneity in the client dataset make it difficult to generalize of the global model aggregated from client models \cite{mendieta2022local}.
A client with a balanced distribution of various locations can learn more diverse mobility patterns, and is therefore helpful for generalization of the global model.
However, in previous federated learning research, clients participating in each round were uniformly sampled\cite{mcmahan2017communication,li2018fedprox}.
Therefore,  we propose the Entropy-Based Sampling (EBS) that allows a client to participate in each round according to the diversity of its mobility pattern.
We derived this diversity of mobility from the entropy of the location distribution. 
In information theory, the entropy of a random variable is the average degree of diversity inherent in the possible outcomes of the variable \cite{shannon2001mathematical}. 

Let $n_{k}$ be the number of data samples in the $k$-th client.
The set of distinct locations appearing in the trajectories $X_k$ is denoted as $L_{k}$.
And the appearing frequency of a specific location $l$ in the $X_k$ can be represented as $n_{k_{l}}$.
Then, the location entropy of the $k$-th client, $E_{k}$, is described as,

\begin{equation}
\label{equation:location_entropy}
E_{k}=-\sum_{l\in L_{k}} \frac{n_{k_{l}}}{n_{k}^{total}} log(\frac{n_{k_{l}}}{n_{k}^{total}}),
\end{equation}
where $n_{k}^{total}$ is the total number of locations appearing in the trajectories $X_k$ of the $k$-th client.
The location entropy $E_{k}$ indicates the degree of diversity of the mobility pattern in the $k$-th client.
The probability to be sampled for the $k$-th client is described as follows:

\begin{equation}
\label{equation:entropy_sampling}
p_{k}=\frac{E_{k}}{\sum_{i=1}^{K} E_{i}}.
\end{equation}
Using the entropy-based sampling, clients with relatively diverse mobility patterns can often participate in every round.

\section{Experiments}

Our experiments aim to address the following research questions:

\noindent\textbf{(RQ1)}: How effective is the FedGeo compared to the existing federated learning frameworks specified in the UNLP task?

\noindent\textbf{(RQ2)}: Does the FedGeo outperform the combination of current state-of-the-art FL frameworks and the UNLP models?

\noindent\textbf{(RQ3)}: How do the different components of FedGeo, such as GAA, LWA, and EBS, affect the performance of UNLP?

\noindent\textbf{(RQ4)}: How do hyperparameters such as local epoch, fraction rate, and self-spatial weight in the GAA affect the performance of the FedGeo model?

\subsection{Datasets}

\;\;\;\;\;Our experiments employed two real-world datasets: (1) mobile signaling data, and (2) GPS trajectory data, denoted as Mobile-T and Geo-Life\cite{zheng2010geolife}, respectively. 
All datasets consist of a set of users' trajectories.
We selected the top ten users based on the number of trajectory records from each dataset and used them in our experiments.
Since both datasets have dense trajectories, unlike other public check-in datasets, they are more suitable for evaluating the UNLP model.

\subsubsection{Mobile-T}
This dataset comprises user trajectories obtained from the mobile operator's base stations, which represents the industrial real-world dataset.
The users' location is recorded by each base station using a signal of the user.
We used the mobility dataset of customers who agreed to collect and analyze their information.
The average density of the base stations in Mobile-T is about 100m, so we pre-processed the location records in Mobile-T into a grid at a 100m scale.
We filtered out location records with an average duration of below five minutes from the Mobile-T.
Then, we extracted trajectories which contain more than ten location records.
We selected the location records at five-minute increments.
The number of distinct locations is 79812.

\subsubsection{Geo-Life}\footnote[1]{https://www.microsoft.com/en-us/download/confirmation.aspx?id=52367}
We used the publicly available GPS trajectory dataset, Geo-Life\cite{zheng2010geolife}, collected by Microsoft Research Asia. 
The trajectories consist of location sequences expressed as GPS coordinates. 
Location records in this dataset were recorded every 1-5 seconds, and we selected location records at a 1-minute increment.
Moreover, we extracted trajectories from over ten location records.
Like Mobile-T, we transformed the location record in this dataset into a grid with a 100m resolution, resulting in 50,008 unique 100m grids. 

    \begin{figure}[h]
    \begin{center}
    \includegraphics[width=1\linewidth]{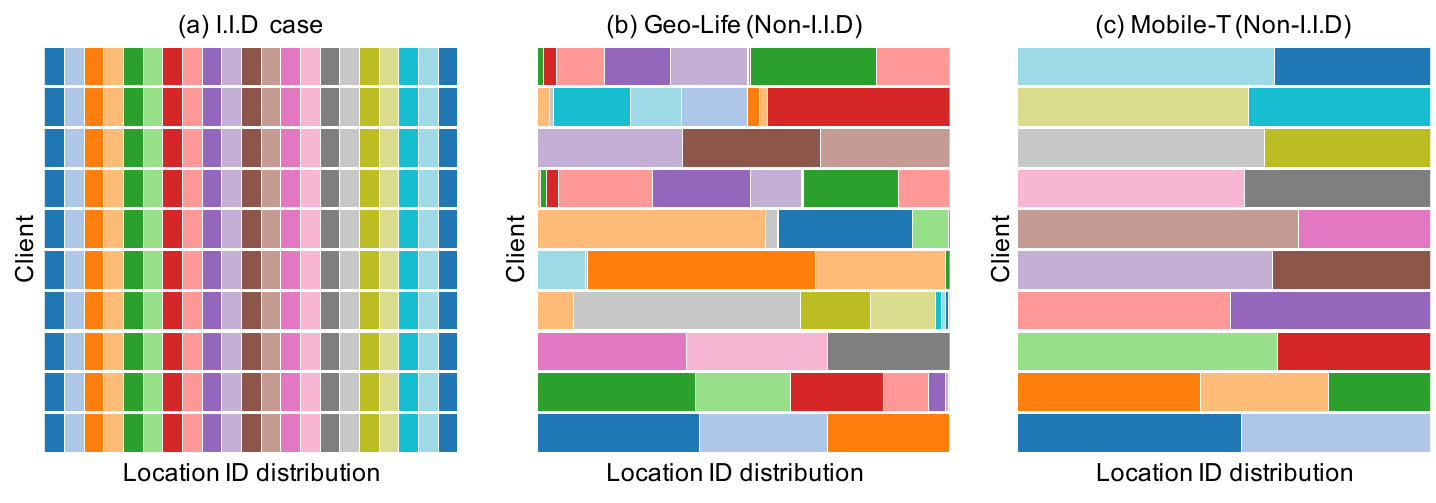}
    \end{center}
    \vspace{-0.5cm}
\caption{
We illustrated the representative 20 location IDs based on the location's frequency in each dataset.
Distribution among location IDs is represented with different colors. 
(a) We described the synthetic distribution of location IDs, ensuring that each client possesses an equal number of location IDs within their respective local datasets.
(b-c) The location ID distributions of Geo-Life and Mobile-T dataset are described.
The heterogeneity in Mobile-T is more pronounced compared to that in Geo-Life.
}
    \label{fig:loc_dist}
    \end{figure}

We measured the degree of the heterogeneity of the two datasets using the Heterogeneity Index (HI) proposed in the \citet{zawad2021curse}.
Considering that fewer classes (i.e., distinct location IDs) per client mean uneven data distribution and large heterogeneity of the dataset, \citet{zawad2021curse} proposed HI as follows:

\begin{equation}
\label{equation:HI}
HI=1-\frac{1}{(C_{max}-1)} \cdot (c-1),
\end{equation}
where $c$ is the maximum number of classes per client, and $C_{max}$ is the total number of classes in the dataset.
Mobile-T and Geo-life have HI of \textbf{0.91} and \textbf{0.72}, respectively, so we can infer that Mobile-T is more heterogeneous data.

This can also be seen by plotting the class distribution for each dataset.
As shown in Figure \ref{fig:loc_dist}, the location ID distribution of each client varies significantly due to diverse user mobility behaviors, resulting in considerable heterogeneity across different users' local data. 
Specifically, the trajectories in Mobile-T (Figure \ref{fig:loc_dist}c) exhibited a considerably imbalanced distribution of locations among all clients.
Geo-Life (Figure \ref{fig:loc_dist}b) also showed the non-IID distribution of trajectories between clients. 
Note that the heterogeneity in Mobile-T is more pronounced compared to that in Geo-Life, indicating that the UNLP task using the federated learning approach is more challenging in Mobile-T and may result in lower overall performance compared to that of Geo-Life.

\begin{table*}[]
\caption{Comparison of UNLP performance with those of previous studies. The top two methods in the federated learning scheme are highlighted in bold and underlined. 
Note that the performance of the centralized learning is an upper bound on the performance of the federated learning.
}
\label{tab:result}
\begin{tabular}{c|cc|cc|cc}
\hline\hline
\multirow{3}{*}{\begin{tabular}[c]{@{}c@{}}\textbf{Model}\\ \textbf{Type}\end{tabular}}                                         & \multicolumn{2}{c|}{\textbf{Dataset}}                                                                                & \multicolumn{2}{c|}{\textbf{Mobilie-T}}                                   & \multicolumn{2}{c}{\textbf{Geo-Life \cite{zheng2010geolife}}}                                      \\ \cline{2-7} 
                                                                                                              & \multicolumn{2}{c|}{\textbf{Metrics}}                                                                                & \multirow{2}{*}{Acc@1}         & \multirow{2}{*}{Acc@5}          & \multirow{2}{*}{Acc@1}          & \multirow{2}{*}{Acc@5}          \\ \cline{2-3}
                                                                                                              & \multicolumn{1}{c|}{\textbf{FL Method}}                                       & \textbf{UNLP model} &                                &                                 &                                 &                                 \\ \hline\hline

\multirow{3}{*}{\begin{tabular}[c]{@{}c@{}}\textbf{Centralized Learning} \\ \textbf{specified in UNLP}\end{tabular}}                              & \multicolumn{1}{c|}{-}                             & DeepMove\cite{mcmahan2017communication}                                    & 9.05$\pm$0.15                           & 30.81$\pm$0.19                            & 17.46$\pm$0.11                            & 35.72$\pm$0.12                           \\ \cline{2-7} 
                                                                                                    & \multicolumn{1}{c|}{-}                             & ST-RNN\cite{liu2016predicting}                                    & 8.23$\pm$0.09                           & 25.08$\pm$0.35                            &   18.58$\pm$0.31                          &  36.46$\pm$0.44                           \\ \cline{2-7}           & \multicolumn{1}{c|}{-}                                  & CTLE\cite{lin2020pre}                                    & 10.71$\pm$0.24                           & 32.39$\pm$1.09                            & 24.69$\pm$0.25                           & 45.12$\pm$0.15                           \\ \cline{2-7} 
                                                                                                                         \hline\hline

\multirow{3}{*}{\begin{tabular}[c]{@{}c@{}}\textbf{Federated Learning} \\ \textbf{specified in UNLP}\end{tabular}}                              & \multicolumn{1}{c|}{PMF\cite{feng2020pmf}}                             & -                                    & 4.12$\pm$0.34                           & 9.39$\pm$2.12                            & 6.19$\pm$0.31                            & 11.38$\pm$0.42                           \\ \cline{2-7} 
                                                                                                    & \multicolumn{1}{c|}{AMF\cite{li2020predicting}}                             & -                                    & 7.81$\pm$1.35                           & 20.66$\pm$4.69                            &   7.99$\pm$0.35                          &  12.37$\pm$0.42                           \\ \cline{2-7}           & \multicolumn{1}{c|}{APF\cite{wang2022location}}                                  & -                                    & 3.88$\pm$0.10                           & 12.31$\pm$0.28                            & 10.51$\pm$0.38                           & 20.66$\pm$0.68                           \\ \cline{2-7} 
                                                                                                              & \multicolumn{1}{c|}{LocationTrails\cite{gurukar2021locationtrails}}         & -                                    & 4.83$\pm$0.47                           & 10.54$\pm$2.44                           & 8.99$\pm$0.32                            & 17.65$\pm$0.34                           \\ \hline
\multirow{15}{*}{\begin{tabular}[c]{@{}c@{}}\textbf{Combination of}\\ \textbf{FL algorithms and} \\ \textbf{UNLP models}\end{tabular}} & \multicolumn{1}{c|}{\multirow{3}{*}{FedAvg\cite{mcmahan2017communication}}} & DeepMove\cite{zhou2018deepmove}                             & 5.28$\pm$0.65                           & 11.40$\pm$2.50                           & 9.25$\pm$0.35                            & 18.66$\pm$0.48                           \\
                                                                                                              & \multicolumn{1}{c|}{}                                                & ST-RNN\cite{liu2016predicting}                               & 3.83$\pm$0.24                           & 9.25$\pm$2.13                            & 6.64$\pm$0.21                            & 12.41$\pm$0.11                           \\
                                                                                                              & \multicolumn{1}{c|}{}                                                & CTLE\cite{lin2020pre}                                  & 5.61$\pm$0.40                           & 9.43$\pm$0.64                           & 8.45$\pm$0.36                           & 14.86$\pm$0.62                        \\ \cline{2-7} 
                                                                                                              & \multicolumn{1}{c|}{\multirow{3}{*}{FedAdam\cite{reddi2020adaptive}}}       & DeepMove\cite{mcmahan2017communication}                             & 4.94$\pm$0.65                           & 10.47$\pm$2.35                           & 9.59$\pm$0.19                            & 20.53$\pm$0.52                           \\
                                                                                                              & \multicolumn{1}{c|}{}                                                & ST-RNN\cite{liu2016predicting}                               & 4.02$\pm$0.54                           & 7.18$\pm$1.32                            & 5.90$\pm$0.29                            & 14.03$\pm$0.42       \\
& \multicolumn{1}{c|}{}                                                & CTLE\cite{lin2020pre}                                  & 6.17$\pm$1.95                           & 19.00$\pm$6.11                     & 8.45$\pm$0.31                           & 14.86$\pm$0.54                           \\ \cline{2-7} 
                                                                                                              & \multicolumn{1}{c|}{\multirow{3}{*}{FedProx\cite{li2018fedprox}}}           & DeepMove\cite{zhou2018deepmove}                             & 5.27$\pm$0.68                           & 12.51$\pm$1.62                           & 9.06$\pm$0.36                            & 18.87$\pm$0.85                           \\
                                                                                                              & \multicolumn{1}{c|}{}                                                & ST-RNN\cite{liu2016predicting}                               & 4.46$\pm$0.37                           & 9.27$\pm$2.12                            & 5.60$\pm$0.19                            & 12.65$\pm$0.39                           \\
                                                                                                              & \multicolumn{1}{c|}{}                                                & CTLE\cite{lin2020pre}                                  & 7.81$\pm$1.59                           & \underline{20.66$\pm$5.14}                           & \underline{13.28$\pm$0.27}                     & \underline{23.28$\pm$0.42}                     \\ \cline{2-7} 
                                                                                                              & \multicolumn{1}{c|}{\multirow{3}{*}{MOON\cite{li2021model}}}                & DeepMove\cite{zhou2018deepmove}                             & 4.48$\pm$0.36                           & 8.25$\pm$0.59                            & 7.90$\pm$0.28                            & 14.41$\pm$0.34                           \\
                                                                                                              & \multicolumn{1}{c|}{}                                                & ST-RNN\cite{liu2016predicting}                               & 4.44$\pm$0.36                           & 8.13$\pm$1.08                            & 6.96$\pm$0.23                            & 12.40$\pm$0.41                           \\
                                                                                                              & \multicolumn{1}{c|}{}                                                & CTLE\cite{lin2020pre}                                  & \underline{8.51$\pm$1.94}                     & 20.21$\pm$5.41                           & 10.86$\pm$0.23                           & 21.87$\pm$0.40                           \\ \cline{2-7} 
                                                                                                              & \multicolumn{1}{c|}{\multirow{3}{*}{\textbf{FedGeo}}}                         & DeepMove\cite{zhou2018deepmove}                             & 5.72$\pm$0.51                           & 13.35$\pm$1.75                           & 9.62$\pm$0.70                            & 19.03$\pm$1.12                           \\
                                                                                                              & \multicolumn{1}{c|}{}                                                & ST-RNN\cite{liu2016predicting}                               & 4.61$\pm$0.37                           & 9.83$\pm$1.79                            & 7.34$\pm$0.27                            & 14.12$\pm$0.57                           \\
                                                                                      & \multicolumn{1}{c|}{}                                                & CTLE\cite{lin2020pre}                                 & \textbf{9.96$\pm$1.49} & \textbf{26.00$\pm$4.43} & \textbf{13.55$\pm$0.42} & \textbf{24.23$\pm$0.63} \\ \hline\hline
\end{tabular}
\end{table*}

\subsection{Experimental Setup}
\subsubsection{Data Partition}
Mobile-T and Geo-Life datasets used in this study were partitioned into ten imbalanced subsets (i.e., clients or users).
We considered each client a user, which is suitable for real-world applications.
We set the  10\% trajectories of each client as the test dataset and evaluated the global model using the union of the test dataset of ten users in each dataset.

\subsubsection{Metrics}
The accuracy of the test dataset was used to measure the performance of the UNLP task.
The \textbf{Acc@\textit{k}} metric determines the proportion of cases where the target location appears in the top $k$ positions (based on the cutoff rate at $k$).
We presented this metric at $k$=1 and $k$=5.

\subsubsection{Baselines}

\noindent \textbf{1) Federated Learning specified in UNLP}
We compared our model with four federated learning frameworks specified in the UNLP task: (1) PMF\cite{feng2020pmf}, (2) AMF\cite{li2020predicting}, (3) APF \cite{wang2022location}, and (4) LocationTrails\cite{gurukar2021locationtrails}.
PMF\cite{feng2020pmf} proposed the long short-term memory (LSTM) based UNLP model with the FedAvg scheme. 
AMF\cite{li2020predicting} aggregated the self-attention based client models with the similarity between the client model and the server model.
APF\cite{wang2022location} combined a global and local UNLP model with a certain ratio to derive a personalized UNLP model.
Lastly, LocationTrails\cite{gurukar2021locationtrails} is a skip-gram-based location embedding model under the FedAvg scheme.

\noindent \textbf{2) Combination of existing FL algorithms and the UNLP models}
We also compared our model with the combination of four federated learning algorithms and the three UNLP models.
We conducted the experiments on following state-of-the-art federated learning frameworks: (1) FedAvg\cite{mcmahan2017communication}, (2) FedAdam\cite{reddi2020adaptive}, (3) FedProx\cite{li2018fedprox}, and (4) MOON\cite{li2021model}. 
In FedProx, we set the parameter of the additional proximal term $\mu$ to 0.5 to derive the best results.
The hyper-parameters $\beta_{1},\beta_{2}$, and $\gamma$ of FedAdam were set to 0.9, 0.995, and $10^{-4}$, respectively.
In addition, we set $\mu$ to $1$ to control the weight of the model-contrastive loss in MOON.

Meanwhile, for the UNLP task on the federated learning, we used three state-of-the-art architectures: (1) DeepMove\cite{zhou2018deepmove}, (2) ST-RNN\cite{liu2016predicting}, and (3) CTLE\cite{lin2020pre}.
In these three models, we commonly employed a 128 dimension embedding layer.
DeepMove and ST-RNN employ an RNN-based model, and CTLE uses a transformer-based network.
For DeepMove and ST-RNN, we used a two-layer RNN encoder with 128 dimensions in our experiment.
In the case of CTLE, we set two stacks of the transformer encoder layers which contained four attention heads with 128 hidden dimensions.
We conducted extensive experiments combining all four of the above federated learning frameworks and the three UNLP models. 

\noindent \textbf{3) Centralized Learning specified in UNLP}
We also measured the performance of the centralized learning method on the three UNLP models presented above (i.e., DeepMove\cite{zhou2018deepmove}, ST-RNN\cite{liu2016predicting}, CTLE\cite{lin2020pre}). 
The centralized learning method naturally outperforms the federated learning method because it collects and trains all the data on a single server. 

\subsubsection{Default Settings} \label{subsubsection:default_setting}
PyTorch and Flower\cite{beutel2020flower}\footnote[2]{https://github.com/adap/flower} were used to implement our model and the other baselines.
In addition, an SGD optimizer with a learning rate of $10^{-4}$ was used for all approaches to train our local models.
The SGD weight decay was set to $10^{-5}$ and the SGD momentum was set to 0.9. 
In our experiments, the length of the input trajectory and the batch size was 32.
We set the number of the federated learning rounds to 50 and 100 for Mobile-T and Geo-life, respectively, which is enough for each dataset to converge.
The number of epochs in each client (i.e., local epochs) is set to ten per round for both datasets.
Then, the fraction rate, the ratio of client selection at each round, is set to 0.2 and 0.4 for both datasets respectively.

\vspace{-0.3cm}
\subsection{Accuracy Comparison (RQ1 and RQ2)} 

\;\;\;\;\;We showed the superiority of our proposed model by comparing it with four federated learning frameworks specified for the UNLP task. 
We reported the best performance on the test dataset and the standard deviation of the last ten rounds of performance in Table \ref{tab:result}.
Our model consistently outperformed these baselines by a significant margin in the $Acc@1$ and $Acc@5$ (RQ1). 
This is because the baseline models cannot account for the heterogeneity of each client's trajectory data.
In particular, although PMF and APF successfully incorporated the LSTM and self-attention module into the UNLP model to capture temporal correlations between locations within a trajectory, they were unable to prevent performance degradation due to the spatial heterogeneity of each client's mobility dataset. 
Similarly, LocationTrails is unable to account for spatial heterogeneity, resulting in poor performance.
In addition, LocationTrails' UNLP model is based on the skip-gram, which can only capture temporal correlations within a limited window size.

Simultaneously, we executed extensive experiments on the Mobile-T and Geo-Life dataset with numerous combinations of the federated learning baselines and UNLP models.
The comparison of the models leads to two conclusions: (1) For most federated learning algorithms, attention-based UNLP models such as CTLE outperform the RNN-based model. 
The RNN-based model focuses on the most recent location when predicting the next location, while the attention-based model adaptively trains all locations to model short- and long-term dependencies between locations \cite{hsieh2019robustness}.
In other words, the RNN-based model intensively trains only a portion of each client's data (i.e., recent locations), resulting in a relatively high degree of data heterogeneity and poor performance.
(2) FedProx and MOON tend to perform better than FedAvg and FedAdam because their models mitigate the heterogeneity of the data, which can alleviate the client drift problem. 
We addressed this issue in the LWA module of FedGeo.

Comparing different federated learning approaches, we observed that our model consistently outperformed other competitive federated learning baselines across both datasets as shown in Table \ref{tab:result}  (RQ2).
Specifically, although the UNLP task on the Mobile-T presents a greater challenge due to its higher degree of heterogeneity compared to Geo-life, our model demonstrated a substantial performance advantage over competing models in the Mobile-T dataset rather than in Geo-life. 
This means that FedGeo is more competitive than other federated learning baselines on the heterogeneous dataset.
As shown in Table \ref{tab:result}, our model showed a higher performance gain Mobile-T with large heterogeneity than Geo-Life, which indicates the robustness of FedGeo. 
In addition, our experiment showed that FedGeo always outperformed other federated learning algorithms for the same UNLP model (Table \ref{tab:result}). 
This is due to three modules of FedGeo effectively handling heterogeneous mobility datasets.
In short, our model was stable and performed well on both datasets, in contrast to all baselines.

\setlength\intextsep{0cm}
\begin{table}[]
\begin{threeparttable}[t]
\caption{Component analysis}
\label{tab:ablation}
\begin{tabular}{l|cc|ll}
\hline \hline
\multicolumn{1}{c|}{\textbf{Dataset}}    & \multicolumn{2}{c|}{\textbf{Mobile-T}} & \multicolumn{2}{c}{\textbf{Geo-Life}} \\ \hline
\multicolumn{1}{c|}{\textbf{Components}} & Acc@1         & Acc@5         & Acc@1         & Acc@5        \\ \hline
(A) FedAvg                         & 5.61          & 9.43         &  8.45        & 14.86              \\ \hline
(B) +GAA                         & 8.16          & 18.70         &  11.22        & 22.02              \\
(C) +LWA                         & 7.34          & 16.90         &  10.76        & 21.34             \\
(D) +EBS                         & 9.15          & 22.33         &  10.35        & 20.26             \\
(E) +GAA+LWA                     &      7.61         &           18.26    & 11.38         & 21.78        \\
(F) +GAA+EBS                     & 9.69          & 23.00         & 11.24         & 22.99        \\
(G) +LWA+EBS                     & 9.18          & 24.16         & 11.12         & 21.84        \\ \hline
(H) \textbf{+GAA+LWA+EBS}                      & \textbf{9.96}          & \textbf{26.00}         & \textbf{13.55}         & \textbf{24.23}        
 \\ \hline \hline
\end{tabular}
\end{threeparttable}
\end{table}

    \begin{figure}[h]
    \begin{center}
    \includegraphics[width=1\linewidth]{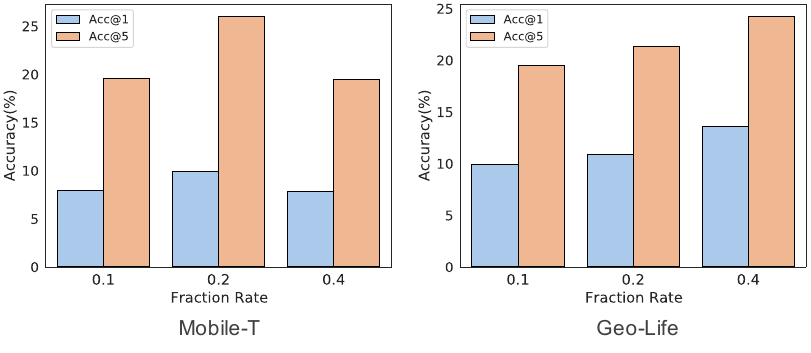}
    \end{center}
\caption{Comparison of UNLP performance for different fraction rates.
}
    \label{fig:ablation_graph_fraction_rate}
    \end{figure}

\subsection{Ablation Study (RQ3)}
\;\;\;\;\;We investigated the effectiveness of each module of FedGeo. 
Table \ref{tab:ablation} shows the experimental results, depending on whether each module was applied or not on FedAvg.
Our component analysis was based on CTLE\cite{lin2020pre} with the default setting described in Section \ref{subsubsection:default_setting}
The models with a single component (B,C,D) generally showed better $Acc@1$ and $Acc@5$ than FedAvg (A).
We found that all three modules (i.e., GAA, LWA, and EBS) contribute to performance improvements when applied to FedAvg.
Rows (E) and (H) showed the effect of the EBS module. 
In both datasets, $Acc@1$ and $Acc@5$ of the model increased by more than $2\%$.
The effect of the EBS module, which selects clients with relatively diverse location data for more frequent participation in the federated learning, was more pronounced in the Mobile-T due to its higher heterogeneity compared to Geo-Life.

The effect of the LWA module can be explained with rows (F) and (H).
A small increment of $Acc@1$ was observed in Mobile-T, but a relatively large increment of about $2\%$ was shown in Geo-Life.
Rows (G) and (H) indicated the effect of the GAA module, which further increased model performance.
The impact of the GAA module on the model performance (e.g., $Acc@1$) was greater than that of the LWA module in both datasets.
Therefore, depending on the dataset's characteristics, an appropriate combination of three modules can lead to improved model performance.

    \begin{figure}
    \begin{center}
    \includegraphics[width=1\linewidth]{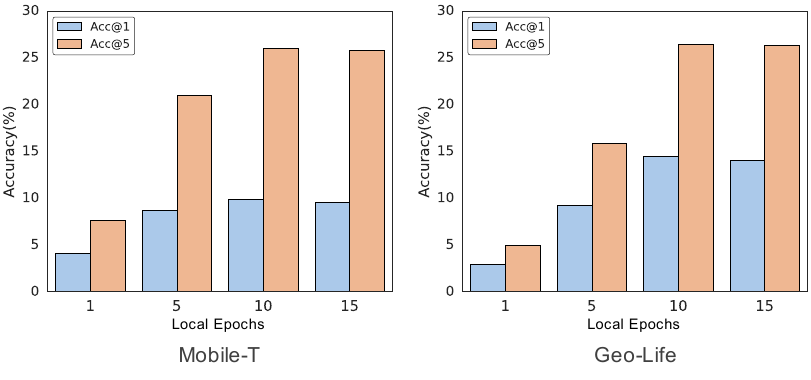}
    \end{center}
\vspace{-0.5cm}
\caption{Comparison of UNLP performance for different local epochs.
}
    \label{fig:ablation_graph_epoch}
    \end{figure}

\subsection{Additional Experiments and Discussions (RQ4)}

\subsubsection{Fraction Rate}
The fraction rate is the ratio of client selection at each round.
As shown in Figure \ref{fig:ablation_graph_fraction_rate}, we demonstrated the effects of the fraction rate in FedGeo using CTLE.
We tuned the fraction rate from \{0.1, 0.2, 0.4\}.
When we set the fraction rate to 0.2 in Mobile-T, our model showed the best performance over the other variants.
Otherwise, the model with a fraction rate of 0.4 showed the best performance in Geo-Life.
This indicates that our model showed reasonable performance even with a small fraction rate.
In the case of Mobile-T, where the data heterogeneity is relatively high, it can be seen that too high a fraction rate has a detrimental effect on performance.

\subsubsection{Number of Local Epochs}
Here, we demonstrated the effects of the number of local epochs in FedGeo using CTLE.
The number of local epochs was set to 1, 5, and 10.
As shown in Figure \ref{fig:ablation_graph_epoch}, our approach is practical after five local epochs.
When the number of epochs is over 10, our model's performance is saturated. 
As a result, a model can plateau or diverge for large numbers of local epochs\cite{mcmahan2017communication}, and we found a moderate number of local epochs (i.e., 10) is necessary for better performance in FedGeo.

\subsubsection{Hyperparameter in GAA ($q$)} \label{subsubsection: gaa_q}
\begin{figure}
    \begin{center}
    \includegraphics[width=1\linewidth]{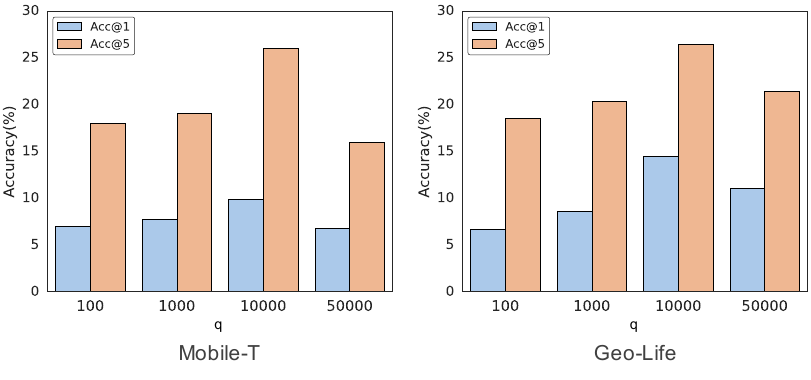}
    \end{center}
\vspace{-0.5cm}
\caption{Comparison of UNLP performance for different $q$.
}
    \label{fig:ablation_graph_q}
    \end{figure}

In the Geographic Adjacency Alignment, diagonal elements of a spatial weight matrix are denoted by $q$ representing the self-spatial weights of a specific location.
The value $q$ represents how much the target location embedding is affected by neighboring locations.
When $q$ is large, neighboring locations have less impact on the target location.
As a result, we found that  $q$ = $10^4$ is a reasonable choice for the best model performance, as shown in Figure \ref{fig:ablation_graph_q}.
This result implies that it is crucial to appropriately determine the number of neighboring locations to maximize the performance of FedGeo, emphasizing the importance of this parameter in the model's overall efficacy.

\subsection{Deployed Solution}
\;\;\;\;\;We validated our model in a real-world application. 
First, we installed the federated learning agent on our customers' mobile phones. 
The federated learning agent contains a UNLP model for each mobile phone and a federated learning module that sends and receives the trained model to and from the server.
As a result, each customer's mobile phone collects the customer's location information (i.e., GPS coordinates) without any leakage to the central server.
Using only this data, each customer's mobile phone trains its own UNLP model and sends only the trained model to the central server. 
The test was conducted for two weeks for 17 customers, and customers who participated in the experiment consented to have their information collected and analyzed.
Then, we used the same test dataset to compare the performance of the UNLP model (i.e., CTLE\cite{lin2020pre}) with the centralized learning and our federated learning method with CTLE\cite{lin2020pre}, as shown in Table \ref{tab:real_world}. 
We found that our model did not show a significant degradation in UNLP performance compared to the centralized learning approach.
Therefore, we demonstrated that our model works reliably in real-world situations.

\begin{table}[]
\caption{Comparison of UNLP performance for real-world applications.}
\label{tab:real_world}
\begin{tabular}{c|ccc}
\hline\hline
\textbf{Method}               & Acc@1 & Acc@3 & Acc@5 \\ \hline
Centralized Learning & 20.21 & 29.11 & 33.91 \\
Federated Learning (Ours)   & 11.52 & 24.08 & 26.33 \\ \hline\hline
\end{tabular}
\end{table}

\section{Conclusion}

\;\;\;\;\;In this study, we propose FedGeo, a privacy-preserving UNLP model with a federated learning framework, which alleviates the heterogeneity of user trajectory data.
FedGeo consists of Geographic Adjacency Alignment to give a prior geographic adjacency relationship to the local client model, Layer-wise Similarity-based Aggregation to handle client drift, and Entropy-based Sampling to allow clients with relatively diverse mobility patterns to participate often in the federated learning.
FedGeo performs better than other federated learning frameworks for model accuracy and performance stability in the UNLP task.

\begin{acks}
This work was supported by the Institute of Information \& communications Technology Planning \& Evaluation (IITP) grant funded by the Korea government (MSIT) (No.2019-0-00075, Artificial Intelligence Graduate School Program (KAIST)), and the National Research Foundation of Korea (NRF) grant funded by the Korea government (MSIT) (No. NRF-2022R1A2B5B02001913).
The authors would like to thank the AI Service Business Division of SK Telecom for providing GPU cluster support.
\end{acks}

\clearpage
\bibliographystyle{ACM-Reference-Format}
\bibliography{ref}

\end{document}